%% file: cmaster.tex
\documentclass[%
reprint,groupedaddress,
amsmath,amssymb,aps,prl]{revtex4-2}
\usepackage[utf8]{inputenc}
\usepackage{graphicx}
\usepackage{dcolumn}
\usepackage{bm}
\usepackage{soul,color}
\usepackage{comment}
\usepackage{lineno}
\usepackage{blindtext}
\usepackage{printlen}
\usepackage[dvipsnames]{xcolor}
\usepackage{titlesec}
\usepackage{upgreek}
\usepackage[normalem]{ulem}
\usepackage{cuted}

\usepackage{bibunits}
\defaultbibliography{references}    
\defaultbibliographystyle{apsrev4-2}     

\titleformat{\subsection}{\bfseries\raggedright}{}{0em}{} 
\titlespacing{\subsection}{0pt}{\baselineskip}{0.5\baselineskip}
\makeatletter
\renewcommand\section{\@startsection{section}{1}{0pt}%
  {16pt} % Space before
  {6pt}  % Space after
  {\normalfont\fontsize{10pt}{12pt}\bfseries\centering}} % Font size and style
\makeatother

\newcommand{\figref}[2]{Fig.\,\textcolor{blue}{#1#2}}
\usepackage[colorlinks=true, linkcolor=blue, citecolor=magenta]{hyperref}

\begin{document}

\twocolumngrid

\begin{bibunit}[apsrev4-2]
{\let\origaddcontentsline\addcontentsline
  \renewcommand{\addcontentsline}[3]{}

\input{cmain_body}

  \putbib
}
\end{bibunit}

\clearpage
\onecolumngrid

\begin{bibunit}[apsrev4-2]

\begin{center}
  \vspace*{1em}
  {\fontsize{12pt}{14pt}\selectfont\bfseries SUPPLEMENTARY INFORMATION}
  \vspace{0.5em}
\end{center}
\tableofcontents

\input{csupp_body}

\putbib
\end{bibunit}

\end{document}

%% file: cmain_body.tex
\title{All-nitride superconducting qubits based on atomic layer deposition}

\author{Danqing Wang$^{1}$, Yufeng Wu$^{1}$, Naomi Pieczulewski$^{2}$, Prachi Garg$^{3}$, Manuel C. C. Pace$^{1}$, C. G. L. Bøttcher$^{4}$, Baishakhi Mazumder$^{3}$, David A. Muller$^{5,6}$, Hong X. Tang$^{1,*}$}

\affiliation{\parbox[t]{\linewidth}{\centering \textcolor{white} {leave a space intentionally}}
\parbox[t]{\linewidth}{\centering \mbox{$^1$Department of Electrical and Computer Engineering, Yale University, New Haven, CT 06511, USA}} 
\parbox[t]{\linewidth}{\centering \mbox{$^2$Department of Materials Science and Engineering, Cornell University, Ithaca, NY 14853, USA}}
\parbox[t]{\linewidth}{\centering \mbox{$^3$ Department of Materials Design and Innovation, University at Buﬀalo-SUNY, Buﬀalo, NY 14260, USA}}
\parbox[t]{\linewidth}{\centering \mbox{$^4$ Department of Applied Physics, Yale University, New Haven, CT 06511, USA}}
\parbox[t]{\linewidth}{\centering \mbox{$^5$School of Applied and Engineering Physics, Cornell University, Ithaca, NY 14853 USA}}
\parbox[t]{\linewidth}{\centering \mbox{$^6$Kavli Institute at Cornell for Nanoscale Science, Cornell University, Ithaca, NY 14853, USA}}
{$^*$hong.tang@yale.edu}}

\begin{abstract}
    The development of large-scale quantum processors benefits from superconducting qubits that can operate at elevated temperatures and be fabricated with scalable, foundry-compatible processes. Atomic layer deposition (ALD) is increasingly being adopted as an industrial standard for thin-film growth, particularly in applications requiring precise control over layer thickness and composition. Here, we report superconducting qubits based on NbN/AlN/NbN trilayers deposited entirely by ALD. By varying the number of ALD cycles used to form the AlN barrier, we achieve Josephson tunneling through barriers of different thicknesses, with critical current density spanning seven orders of magnitude, demonstrating the uniformity and versatility of the process. Owing to the high critical temperature of NbN, transmon qubits based on these all-nitride trilayers exhibit microsecond-scale relaxation times, even at temperatures above 300 mK. These results establish ALD as a viable low-temperature deposition technique for superconducting quantum circuits and position all-nitride ALD qubits as a promising platform for operation at elevated temperatures.
\end{abstract}

\maketitle	

%\linenumbers
%\setlength\linenumbersep{5pt}

\section{Introduction}

Superconducting qubits based on Josephson junctions have emerged as a leading platform for scalable quantum computing over the past few decades \cite{Nakamura1999-la, Koch2007-pu, Siddiqi2021-ub, Google-Quantum-AI-and-Collaborators2025-ct}. Among these, aluminum (Al)-based junctions are the most established, recently achieving wafer-scale fabrication suitable for large-scale processors \cite{Van-Damme2024-mj}. However, such qubits typically operate below 100\,mK. As the number of qubits increases, the cumulative heat load from qubit manipulation imposes stringent demands on cooling power. Raising cryostat temperatures is an efficient way to increase cooling power, motivating the development of qubits that can operate at higher temperatures \cite{Krinner2019-ps, Martin2022-ge, Yang2020-mp}.

Superconductors with higher critical temperatures (T$_\text{c}$) offer a path toward elevated-temperature operation. For instance, niobium (Nb)/Al/AlO$_x$ qubits have demonstrated remarkable performance at elevated temperatures compared to pure Al due to the proximity effect \cite{Anferov2024-jv, Anferov2024-da, PRXQuantum.6.020336}. Epitaxial NbN/AlN/NbN trilayer qubits, fabricated via dc-magnetron sputtering, have also emerged as a promising candidate by leveraging the even higher T$_\text{c}$ of NbN and the potentially low-defect, crystalline AlN tunnel barriers \cite{Nakamura2011-sd, Kim2021-sl}. However, precise control of AlN thickness below 2\,nm remains a key challenge, as small deviations can cause exponential variations in Josephson energy \cite{Tinkham}. Additionally, the quality of the sputtered films is sensitive to substrate and buffer conditions, limiting reproducibility across wafers \cite{Wang2013-hl, Makise2016-nz, Qiu2020-qz}.

Atomic layer deposition (ALD) offers a compelling alternative for fabricating nitride-based superconducting circuits, particularly for its angstrom-level thickness control, excellent uniformity, and broad substrate compatibility \cite{George2010-sq, Deyu2025-yh}. ALD is already integrated into advanced semiconductor manufacturing, enabling conformal coatings of high-$\kappa$ dielectrics in transistor structures and memory cells, seed layers for interconnects, and pin-hole-free encapsulations \cite{Zhao2019-ch, Sheng2018-wc, Waechtler2011-wv, Chen2024-ia}. Its scalability and CMOS compatibility make it a natural candidate for extending into superconducting quantum technologies. In recent years, ALD has been used to deposit high-quality NbN films with T$_\text{c}$ around 13\,K \cite{Sowa2017-qt, Cheng2019-qs}, and Al$_2$O$_3$ tunnel barriers for Josephson junctions \cite{Wilt2017-po, Jhabvala2020-cl}. However, to date, no superconducting qubits have been demonstrated based on ALD-grown trilayer materials. 

In this work, we present the realization of NbN/AlN/NbN superconducting qubits using plasma-enhanced atomic layer deposition (PEALD). These all-nitride qubits are fabricated by a flip-chip bonding technique, which eliminates the use of lossy SiO$_2$ spacers while enabling independent optimization of the junction layers \cite{Anferov2024-jv, Kim2021-sl, Rosenberg2017-vw}. Importantly, the layer thickness can be precisely controlled at the atomic scale by adjusting the number of ALD cycles. DC transport measurements across wafers with varying AlN barrier thicknesses show that the critical current density ($J_\text{c}$) can be tuned over seven orders of magnitude, demonstrating the scalability and reproducibility of the ALD process. Transmon qubits fabricated from these trilayers exhibit microsecond-scale relaxation times, even when measured at temperatures above 300\,mK. These results establish ALD-grown nitride trilayers as a promising material platform for scalable superconducting qubits with potential for elevated-temperature operation.

\begin{figure*}[t]
    \centering
    \includegraphics [width=1\linewidth]
    {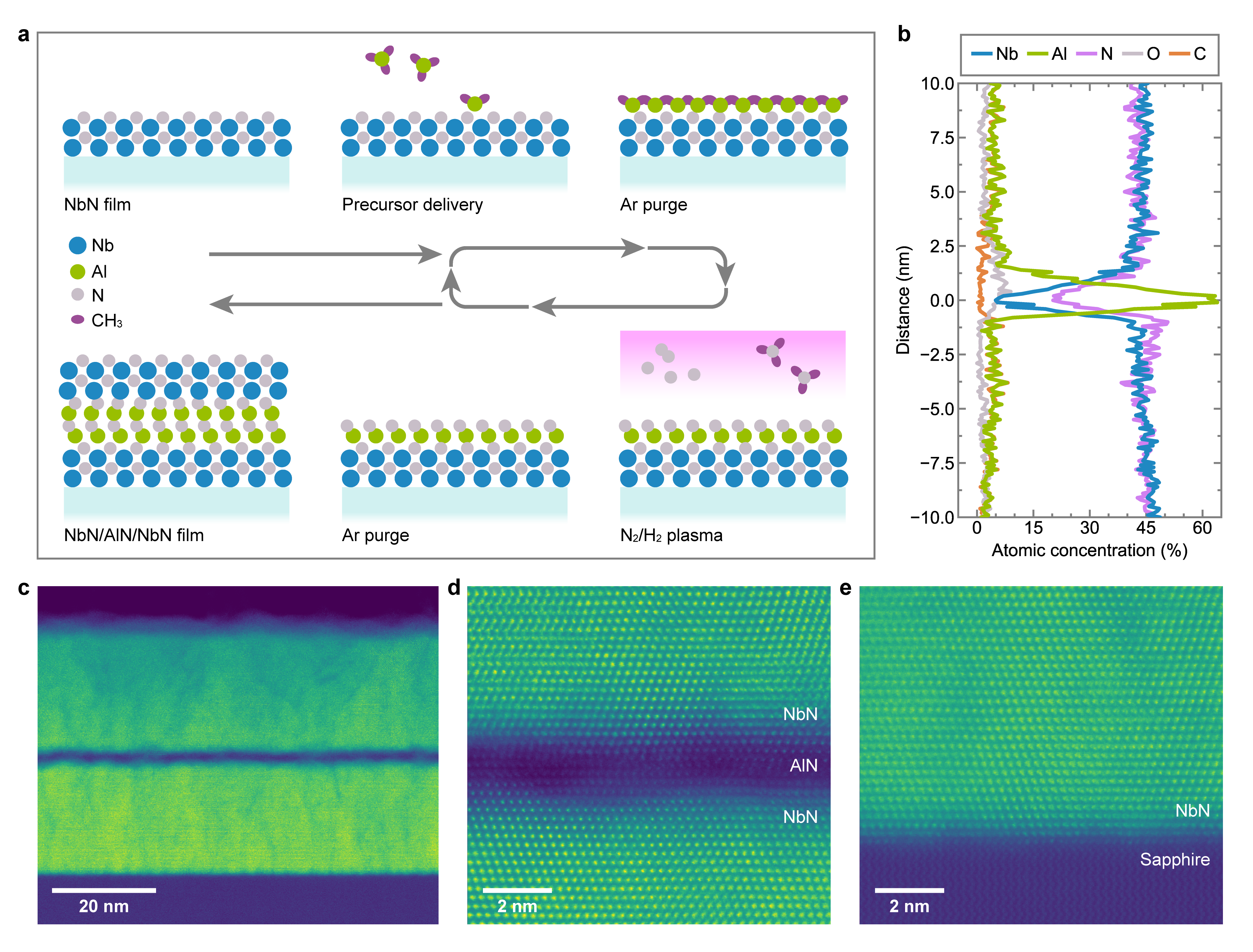}
    \caption{\textbf{ALD growth sequence and atomic-scale characterization of NbN/AlN/NbN trilayer.} \textbf{a}, ALD process flow for trilayer deposition, following the arrows. NbN is first deposited on a c-plane sapphire substrate using the precursor TBTDEN. The Al-containing precursor (TMA) is then introduced and purged, followed by plasma-enhanced nitridation that converts surface-adsorbed Al into AlN. Residual ions and byproducts are subsequently removed in a second purge. Repeated cycles form the AlN barrier and NbN top layer, completing the NbN/AlN/NbN trilayer stack. \textbf{b}, 1D atomic concentration profile extracted from APT of a 15\,nm-diameter cylindrical volume. The y-axis origin (0\,nm) corresponds to the center of the AlN barrier; negative and positive distances denote the bottom and top NbN layers, respectively. \textbf{c}, Overview cross-sectional STEM image. \textbf{d}, Zoom-in STEM image of NbN/AlN/NbN interfaces. \textbf{e}, Magnified STEM image of NbN/sapphire interface.}
    \label{fig:1}
\end{figure*}

\begin{figure*}[t]
    \centering
    \includegraphics [width=1\linewidth]
    {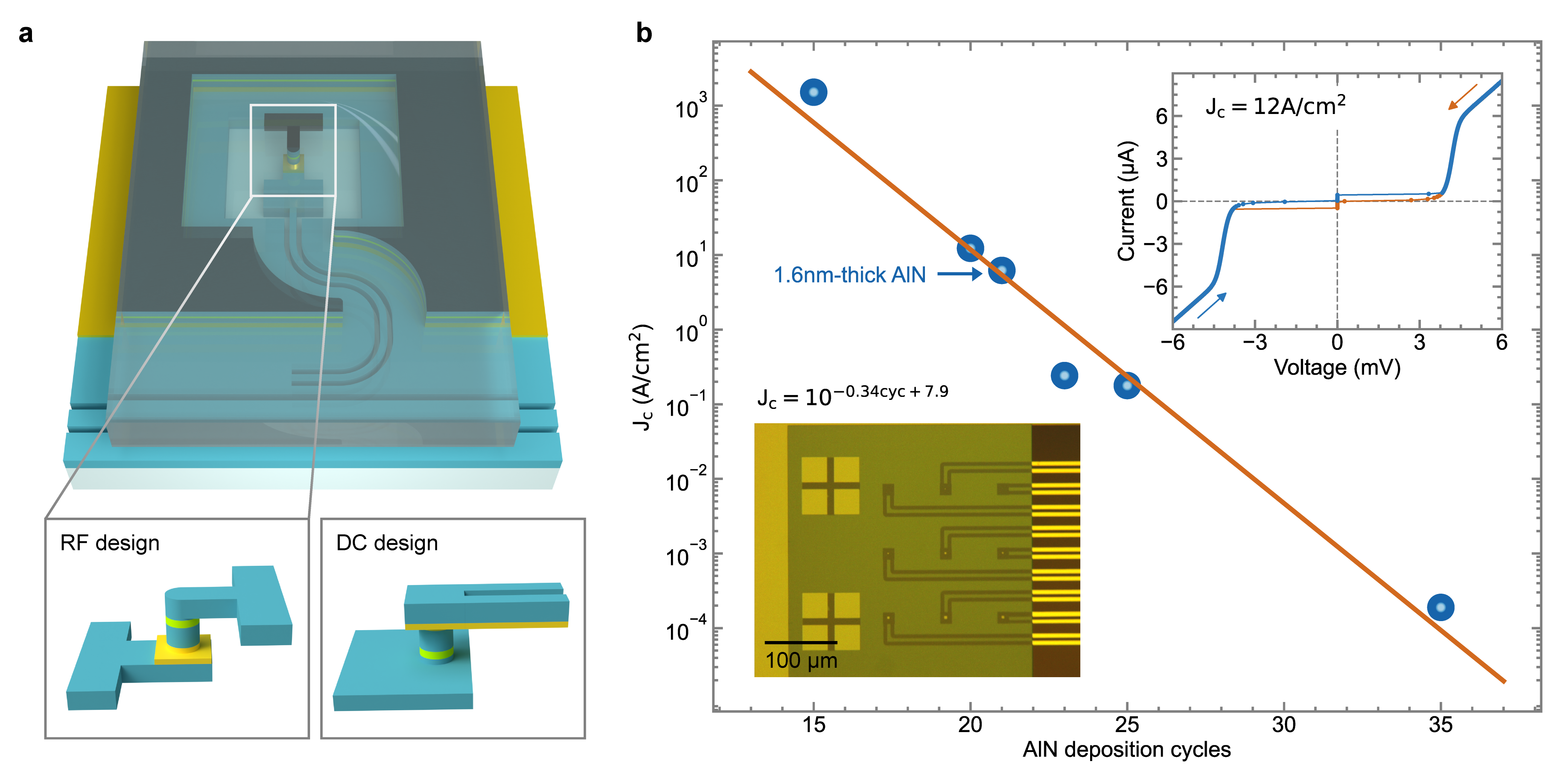}
    \caption{\textbf{Device schematics and IV characterization at 4\,K.} \textbf{a}, A flip-chip qubit model, where the C-chip, consisting of a readout resonator and a transmission line, is at the bottom, while the Q-chip, hosting a junction, is on top. Blue, green, and yellow represent NbN, AlN, and gold, respectively. The lower-left panel shows a zoom-in model of the junction area, where two shunting pads ensure strong coupling of a qubit to the readout resonator. The lower-right panel shows a close-up view of a device used for DC characterization, in which the unetched bottom NbN film serves as the ground plane and the junction top electrode is flip-chip bonded to NbN-gold wires for readout. \textbf{b}, Dependence of $J_\text{c}$ on the number of AlN deposition cycles. Blue dots represent experimental data from wafers grown over a year, while the orange curve is an exponential fit. The arrow indicates that 21 ALD cycles deposit $\sim$1.6\,nm-thick AlN, as estimated by STEM. The inset microscopic image shows nine junctions with two alignment crosses for flip-chip bonding. The inset IV curves correspond to a junction with 7\,$\mathrm{\upmu m}$ diameter.}
    \label{fig:2}
\end{figure*}

\begin{figure*}[t]
    \centering
    \includegraphics [width=1\linewidth]
    {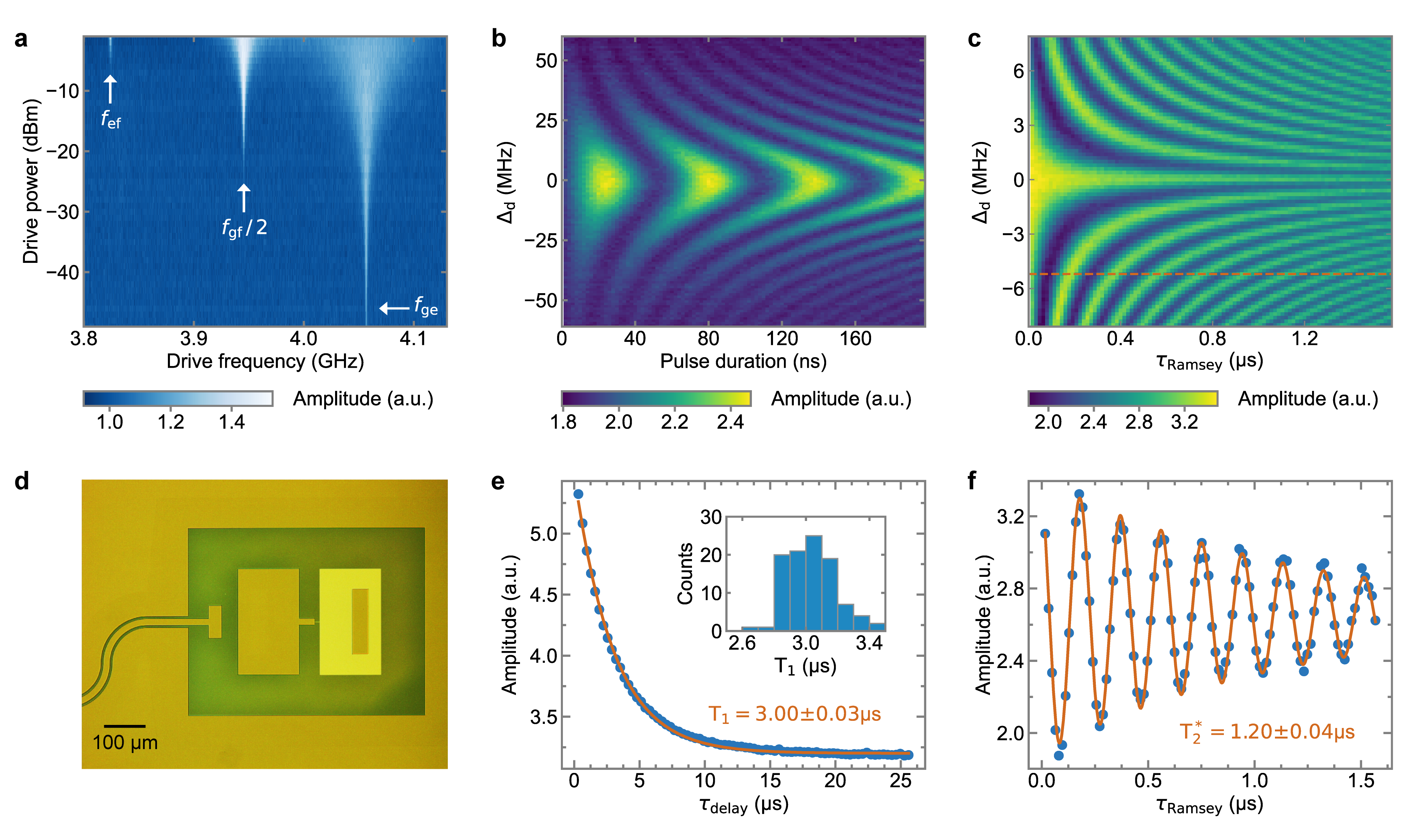}
    \caption{\textbf{Qubit performance at the base temperature.} \textbf{a}, Power-dependent qubit spectroscopy. The power refers to the output level of a signal generator. The color bar represents the voltage amplitude of the transmitted probe tone. \textbf{b}, Chevron plot of Rabi oscillations. \textbf{c}, Ramsey oscillations measured near $f_\text{q}$ with varying $\Delta_\text{d}$. \textbf{d}, Microscopic image of a qubit and part of its readout resonator. \textbf{e}, $T_1$ characterization. The inset histogram summarizes the $T_1$ values from 100 repeated measurements taken over a two-hour period. The blue dots in the main graph represent averaged data from these repeated measurements. The orange curve is an exponential fit, yielding an average $T_1$ of 3.0\,$\mathrm{\upmu s}$. \textbf{f}, Extracted Ramsey oscillation from \textbf{c}, at $\Delta_\text{d}$\,=\,–5.2\,MHz. Orange curve is from a fitting model: $A_0+Ae^{-\tau/T_2^*}\mathrm{cos}(2\pi\Delta_\text{d}\tau+\phi_0)$.}
    \label{fig:3}
\end{figure*}

\begin{figure}[b]
    \centering
    \includegraphics [width=1\columnwidth]
    {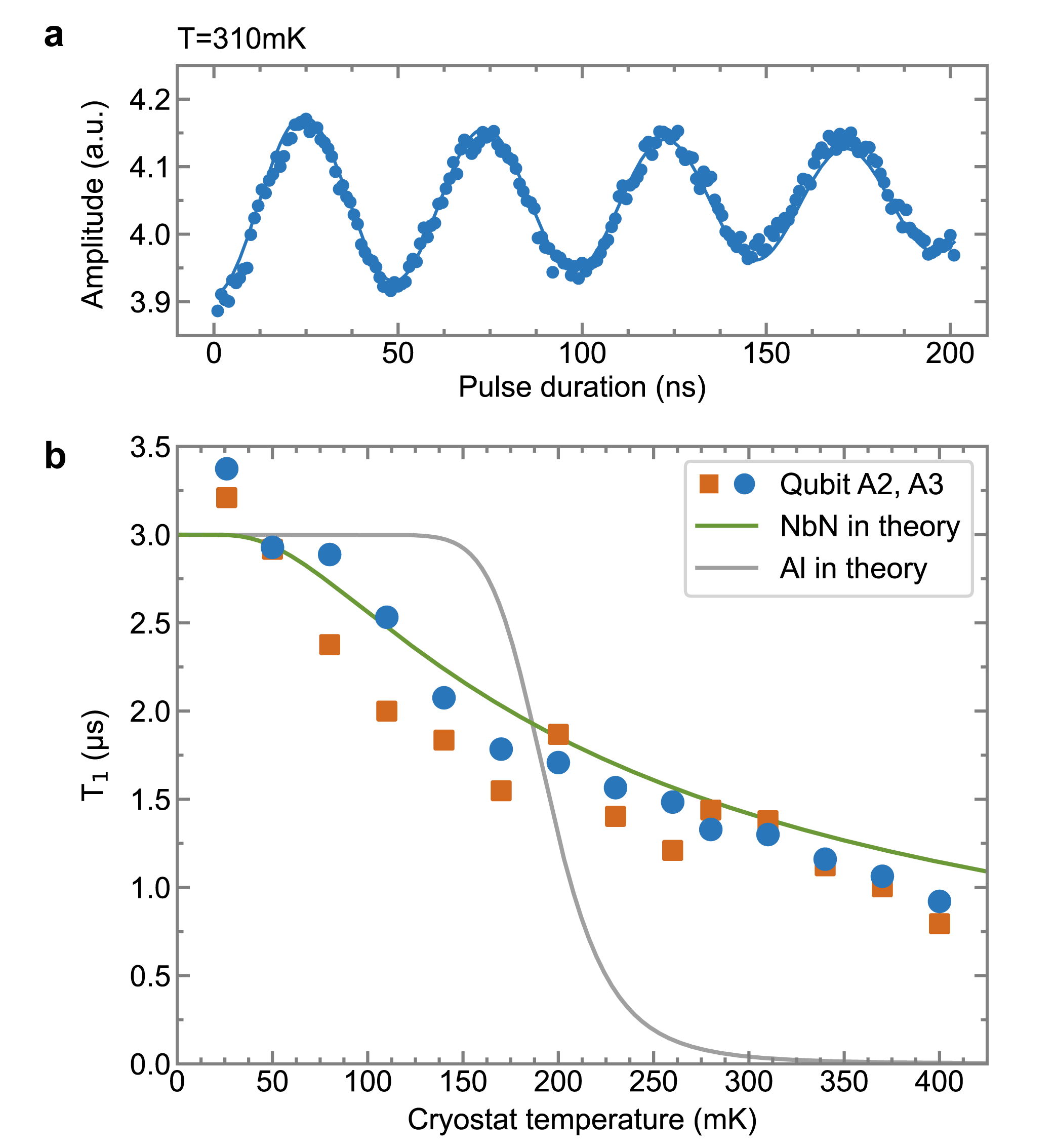}
    \caption{\textbf{Qubit performance at elevated temperatures.} \textbf{a}, Rabi oscillation of qubit A3 as a function of pulse duration at 310\,mK. \textbf{b}, orange squares and blue dots represent the $T_1$ values of qubits A2 and A3, respectively, across a temperature sweep. The green and gray curves show the theoretical temperature dependence of $T_1$, as anticipated by the spin-boson model for NbN-based qubits and the quasiparticle relaxation model for Al-based qubits, respectively, both assuming $T_1$\,=\,3\,$\mathrm{\upmu s}$ at 10\,mK.}
    \label{fig:4}
\end{figure}

\section{Results}
\subsection{ALD growth of NbN/AlN/NbN films and materials characterization}

NbN/AlN/NbN trilayers are deposited on c-plane sapphire substrates using a remote plasma ALD system at temperatures ranging from 300\,$^{\circ}$C to 400\,$^{\circ}$C. Here, sapphire is used for its optical transparency, which facilitates flip-chip alignment. Similar trilayer growth is also achievable on silicon, supporting future CMOS-compatible integration with improved alignment strategies. As illustrated in the process schematic (\figref{\ref{fig:1}}{a}), layer-by-layer deposition of the nitride stack is achieved via alternating ALD cycles \cite{Sowa2017-qt, Alevli2011-zo, Goerke2015-qn}. Each cycle begins with the introduction of metal precursors (TBTDEN for NbN and TMA for AlN), carried by argon (Ar), followed by an Ar purge to ensure uniform adsorption on the substrate surface. A subsequent plasma step using a nitrogen/hydrogen ($\mathrm{N_2/H_2}$) mixture initiates metal nitridation, after which a second Ar purge removes residual ions and byproducts. This cycle repeats until the desired thickness is achieved, forming a NbN/AlN/NbN trilayer. This self-limiting surface chemistry highlights ALD's capability of precise thickness control. To promote interface clarity and minimize residual contamination, additional ten cycles of plasma treatment are applied at the NbN/AlN interfaces. 

To assess the structural quality of the nitride trilayers, cross-sectional scanning transmission electron microscopy (STEM) is performed. The wider field-of-view image in \figref{\ref{fig:1}}{c} shows a uniform trilayer with a $\sim$1.6\,nm AlN barrier. The high-magnification image (\figref{\ref{fig:1}}{e}) reveals cubic-phase NbN deposited on the c-plane sapphire substrate, consistent with previous reports \cite{Shibalov2023-pi, Dang2021-jp}. The NbN layers consist of $\sim$10\,nm-wide twin domains oriented along the [111] direction, separated by vertical grain boundaries (Supplementary Section I). The atomic-resolution image of the NbN/AlN interfaces (\figref{\ref{fig:1}}{d}) reveals atomically sharp interfaces. Interestingly, the first few monolayers of AlN frequently exhibit features resembling the metastable cubic phase \cite{Yaddanapudi2018-cp, Chen2019-dy}, which then transitions into the thermodynamically favored hexagonal structure in subsequent layers \cite{Shih2017-zc}. While the detailed phase structure of AlN warrants further investigation, the well-defined trilayer geometry indicates its suitability for high-quality Josephson junctions. A $\sim$2\,nm amorphized NbO$_x$ layer is also observed on the top surface of the layer stack, attributed to air exposure, but this thin oxide can be removed during subsequent device fabrication.  

Atomic-scale chemical composition of the trilayers is further analyzed using atom probe tomography (APT). A representative 1D concentration profile, extracted from a 15\,nm-diameter cylinder cut across the layer stack, is shown in \figref{\ref{fig:1}}{b}. The Nb and Al profiles confirm the presence of a well-defined AlN barrier with minimal interdiffusion, while the N profile verifies successful nitridation across the interfaces. The elevated oxygen concentration ($\sim$5\,\%) in the AlN layer, together with detectable carbon residues within all layers, suggests opportunities for further optimization of the growth process.

\subsection{Josephson junction fabrication and DC transport characterization} 
The atomic-layer-deposited nitride trilayer offers a potentially high-quality AlN barrier with reduced oxygen content compared to the amorphous AlO$_x$ barrier in Al-based qubits \cite{Wang2013-hl,Kim2021-sl}. However, forming Josephson junctions with this trilayer requires a more intricate fabrication process to establish a wiring layer that connects the top electrode while maintaining isolation from the bottom electrode. A conventional approach employs dielectric sidewall spacers for isolation \cite{Nakamura2011-sd,Qiu2020-qz,Anferov2024-jv, Gronberg2017-fa}, but these can introduce additional microwave loss which degrades qubit coherence, and the added fabrication complexity may create unintended lossy byproducts. To avoid dielectric spacers and simplify processing, we implement a flip-chip technique. Josesphson junctions are patterned on the NbN/AlN/NbN trilayer chip (Q-chip), while a separate NbN single-layer chip (C-chip) defines the readout wiring. This modular fabrication approach offers greater flexibility, with the two chips interconnected via gold-gold bonding \cite{Higurashi2017-sc}. As illustrated in \figref{\ref{fig:2}}{a}, the 3D structure in the lower-right panel enables junction transport measurements through galvanic connections, while the other panels depict the flip-chip qubit realization  with capacitive coupling to the environment (details in a later section).

The Josephson effect is characterized through DC current-voltage (IV) measurements at 4\,K, using a typical device layout shown in the lower-left panel of \figref{\ref{fig:2}}{b}. As illustrated in the upper-right panel, the junctions display hysteretic behavior under bidirectional current sweeps. In the forward sweep (blue curve) with a positive current bias, a supercurrent flows up to the switching current ($I_\text{sw}$), beyond which Cooper pairs break, leading to a finite voltage readout. The observed $I_\text{sw}$ is lower than the theoretical critical current ($I_\text{c}$), likely due to environmental magnetic fields and residual RF noise. Above $I_\text{sw}$, quasiparticle-assisted tunneling dominates until the junction reaches the gap current ($I_\text{g}$), transitioning to the normal state. The reverse sweep (orange curve) retraces this behavior, with a return to the subgap regime as the current is reduced. We estimate $I_\text{c}$ using an empirical relation $I_\text{c}$\,=\,$\pi I_\text{g} / 4$, and determine the superconducting energy gap ($2\Delta$) from the gap voltage ($V_\text{g}$), taken at a bias current $I_\text{g}/2$ \cite{Tinkham, Wang2013-hl}. A key figure of merit for junction quality is the ratio of subgap resistance ($R_\text{sg}$) to normal-state resistance ($R_\text{n}$). The junction discussed here exhibits $R_\text{sg}/R_\text{n}$\,=\,47 and $V_\text{g}$\,=\,4.2\,mV. Across all measured wafers, the best device reaches $R_\text{sg}/R_\text{n}$\,=\,55 and $V_\text{g}$\,=\,4.3\,mV (Supplementary Section IV), values that remain below but comparable to epitaxial NbN/AlN/NbN junctions \cite{Wang2013-hl, Qiu2020-qz}.

For a structurally uniform and pinhole-free tunnel barrier, the room-temperature conductance of Josephson junctions is expected to scale linearly with the junction area. This behavior is confirmed by our measurements (Supplementary Section IV), which reveal an inverse relationship between resistance and area across an array of junctions on a single chip, indicating high lateral uniformity of the ALD-grown films. To further assess its ability of thickness control, we characterize chips from multiple wafers with varying AlN deposition cycles to extract $J_\text{c}$ values at 4\,K. As shown in \figref{\ref{fig:2}}{b}, $J_\text{c}$ which spans a wide range from $\mathrm{10^{-4}}$ to $\mathrm{10^{3}}\,\mathrm{A/cm^2}$ and decreases exponentially with increasing AlN cycles, following a relation $J_\text{c}$\,$\propto$\,$10^{-0.34\, \mathrm{cycle}}$. This strong correlation demonstrates that the AlN thickness increases linearly with ALD cycle number, even at the angstrom scale.

\subsection{Qubit realization and characterization}
Having established the Josephson tunneling characteristics of ALD-grown NbN/AlN/NbN trilayers, we next demonstrate transmon qubits based on this material platform. As illustrated in the flip-chip schematic (\figref{\ref{fig:2}}{a}) and the optical micrograph of a fabricated device (\figref{\ref{fig:3}}{d}), the Q-chip hosts junctions shunted by capacitive pads, while the C-chip contains the complementary pads and readout resonators. The Q-chips are fabricated from two wafers with $J_{\mathrm{c}}$ of 6 and 0.8\,$\mathrm{A/cm^2}$; together with the variation in junction sizes, the resulting junction energy participation ratios ($p_{\mathrm{J}}$) range from 0.2 to 0.8 (Supplementary Section V). Notably, qubits with the highest $p_{\mathrm{J}}$ resemble ``\textit{mergemons}"---a transmon variant \cite{Zhao2020-gg,Mamin2021-hk}---featuring micrometer-scale junctions and compact footprints, which are advantageous for large-scale quantum processors. The smallest junction diameter used in this study is 0.8\,$\mathrm{\upmu m}$, demonstrating compatibility with photolithographic patterning and potential for integration with industrial fabrication workflows. To highlight its atomic layer deposition origin, we hereafter refer to this qubit structure as the \textit{ALDmon}.

Similar to conventional transmons \cite{Blais2021-dh, Place2021-fg}, the \textit{ALDmons} are designed with transition frequencies ($f_\text{q}$) of 4--5\,GHz and are capacitively coupled to quarter-wave resonators operating in the 6--7\,GHz range ($f_\text{c}$), with coupling strengths ($g/2\mathrm{\pi}$) of approximately 50--70\,MHz. Measurements are conducted in a dilution refrigerator at a base temperature of 10\,mK. Qubit characterization begins with two-tone spectroscopy, where the transmission of a weak probe tone fixed at $f_\text{c}$ is continuously monitored and a qubit drive tone is swept in power and frequency. As shown in \figref{\ref{fig:3}}{a}, at low drive power, the qubit transitions from ground to the first excited state ($f_\text{ge}$, denoted $f_\text{q}$ in this study), observed as a change in the transmission amplitude of the probe tone. As the drive power increases, higher-energy transitions ($f_\text{gf}/2$ and $f_\text{ef}$) become resolvable and the anharmonicity ($\alpha/2\mathrm{\pi}$) of 223\,MHz can be extracted. The clean 2D spectroscopy map, free from spurious transitions, points out the high quality of the \textit{ALDmon}'s tunnel barrier.

Based on the established qubit spectroscopy, we investigate qubit dynamics in the time domain \cite{Krantz2019-cb, Gao2021-oq}. The detailed measurement setup is described in Supplementary Section III, where time-domain pulses at a drive frequency ($f_\text{d}$) and at $f_\text{c}$ are used to excite the qubit and read out its state, respectively. As shown in \figref{\ref{fig:3}}{b}, at zero detuning, the qubit exhibits coherent oscillations between the ground $|g\rangle$ and the excited state $|e\rangle$ as the drive pulse duration increases, with a Rabi frequency $\Omega_\text{R}$ of 18\,MHz. The brightest fringes on the 2D map correspond to the maximum $|e\rangle$ population. As detuning ($\Delta_\text{d}$\,=\,$f_\text{d}-f_\text{q}$) increases, the oscillation frequency rises to $\Omega$\,=\,$\sqrt{\Omega_\text{R}^2+\Delta_\text{d}^2}$, while the excitation amplitude decreases. The resulting chevron pattern enables extracting the $\pi$-pulse duration at each detuning. Qubit relaxation time ($T_1$) is then measured by sweeping the delay ($\tau_\text{delay}$) between the $\pi$-pulse excitation and the recording of the probe transmission, producing the result shown in \figref{\ref{fig:3}}{e}. A $T_1$ of 3.00\,$\pm\,0.03\,\mathrm{\upmu s}$ is obtained from 100 repeated measurements conducted over a two-hour period, with the statistical distribution displayed in the inset of \figref{\ref{fig:3}}{e}.

With the calibrated Rabi pulses, Ramsey measurements are performed to obtain the qubit coherence time ($T_2^*$). The qubit is first driven to the $(|g\rangle+|e\rangle)/\sqrt{2}$ state by a $\pi/2$-pulse, then allowed to evolve freely for varying time intervals ($\tau_{\mathrm{Ramsey}}$), followed by a second $\pi/2$-pulse and a final readout pulse. As shown in \figref{\ref{fig:3}}{c}, when $\Delta_\text{d}$\,=\,0\,MHz, the exponential decay of the readout amplitude reflects the loss of qubit coherence. When driven by a slightly detuned $\pi/2$-pulse, the qubit accumulates a phase difference between $|g\rangle$ and $|e\rangle$ during its free evolution, leading to Ramsey oscillations in the time domain with frequency $\Delta_\text{d}$. By fitting the exponential decay envelopes, we extract a $T_2^*$ of 1.20\,$\pm$\,0.04\,$\mathrm{\upmu s}$, as shown in \figref{\ref{fig:3}}{f}, which is on the same order as $T_1$.

A distinctive advantage of nitride superconductors is their high $T_\text{c}$, which makes it feasible to operate all-nitride qubits at elevated temperatures where Al-based qubits suffer from thermally excited quasiparticles \cite{Paik2011-kg, PhysRevB.84.064517, Serniak2018-rz}. To assess this capability, the performance of \textit{ALDmons} in temperature regimes beyond the reach of conventional transmons is investigated. As a demonstration, a Rabi experiment is conducted at 310\,mK on qubit A3 (previously discussed) and reveals well-resolved oscillations between $|g\rangle$ and $|e\rangle$ (\figref{\ref{fig:4}}{a}). This result confirms that NbN-based qubits can function at higher temperatures, despite the reduced signal to noise ratio caused by thermal photon population \cite{Jin2015-na}. To further investigate temperature dependence, we conduct systematic $T_1$ measurements on qubits A2 (properties in Supplementary Section V) and A3 over a five-day temperature sweep from 400\,mK down to 26\,mK. At each temperature, qubits are allowed to thermalize for at least two hours prior to data acquisition. As shown in \figref{\ref{fig:4}}{b}, the measured gradual decrease of $T_1$ with temperature agrees well with the predictions from the spin-boson model, $T_1$\,$\propto$\,$\left(1+\coth\left(\frac{2\pi\hbar f_\text{q}}{2k_\text{B}T}\right)\right)^{-1}$ \cite{Lisenfeld2007-tr}, indicated by the green curve. Crucially, unlike Al-based devices modeled by the gray curve in \figref{\ref{fig:4}}{b}, where $T_1$ exhibits a sharp drop beyond 200\,mK \cite{PhysRevB.84.064517, PhysRevApplied.23.054079}, our NbN/AlN/NbN \textit{ALDmons} offer a significant operational advantage for future quantum technologies requiring relaxed thermal constraints.

\section{Discussion}
In conclusion, we have demonstrated a new material platform for superconducting qubits based on atomic-layer-deposited NbN/AlN/NbN. Atomic-scale structural and compositional analyses confirm the uniformity and interface sharpness of the AlN tunnel barriers, while transport measurement highlights the precise thickness control inherent to the ALD process. The ability to span $J_\text{c}$ over seven orders of magnitude validates the versatility of these trilayers for broad applications. In particular, qubits have been developed through integration with a flip-chip architecture. The $ALDmons$ operate reliably not only at conventional millikelvin temperatures but also above 300\,mK, expanding the temperature window for superconducting quantum devices.

Experimentally, $ALDmons$ exhibit relaxation times in the range of 1--4\,$\upmu$s (Supplementary Section V), corresponding to quality factors ($Q_\text{q}$\,=\,$2\pi f_\text{q} T_1$) of 0.4--0.9$\times 10^5$, motivating further investigation into potential loss mechanisms. As detailed in Supplementary Section VI, simulations indicate that the finite resistance of the gold bonding layer contributes a quality factor exceeding $3\times10^6$, while the subgap conduction path through the AlN barrier corresponds to a $Q$ above $5\times10^5$. These estimations indicate that neither mechanism is the dominant source of loss in our current devices. Combined analysis and comparison with experimental data further confirm that the potential piezoelectricity of the AlN barrier does not limit the present $Q_\text{q}$. Although the predominant loss mechanisms in $ALDmon$ remain unresolved, improvements in substrate quality, NbN dry etching, and device packaging are expected to further enhance qubit coherence \cite{Anferov2024-da, Ganjam2024-bc, Huang2021-co}.

With continued process optimization, \textit{ALDmons}, potentially extending beyond the demonstrated NbN/AlN system to other nitride combinations, would be well-positioned to explore high-frequency operation in the millimeter wave regime and robust performance at elevated temperatures. These capabilities pave the way for a scalable and industry-compatible nitride-based platform for next-generation superconducting quantum technologies.

\section{Methods}
The ALD process was carried out using a Ultratech/Cambridge Fiji G2 plasma system. Sapphire substrates were loaded into the chamber after sequential cleaning with organic solvents (NMP, acetone, and IPA), followed by a 10 min piranha treatment. Prior to deposition, the substrates underwent one cycle of $\mathrm{N_2}$ and $\mathrm{H_2}$ plasma surface activation. Tert-butylimido tris(diethylamido)niobium (TBTDEN) was the precursor for NbN deposition, while trimethylaluminum (TMA) served as the precursor for AlN. The plasma-enhanced deposition process utilized $\mathrm{N_2}$ and $\mathrm{H_2}$ as reactive gases. Across all wafers described in the main text, the two NbN layers in the trilayer structure maintained equal thickness, ranging from 25\,nm to 50\,nm.

A cross-section lamella for STEM was prepared using a Thermo Fisher Helios G4 UX Focused Ion Beam (FIB). Protective carbon and platinum layers were deposited on the lamella and prepared with a final milling step of 2\,keV to reduce damage. STEM measurements were taken with an aberration-corrected Thermo Fisher Spectra 300 CFEG operated at 300 keV.

Needle-shaped APT specimens were prepared using a Thermo Fisher Helios 5UC dual-beam FIB/scanning electron microscope (SEM). The sharp APT needles were prepared by standard site-specific lift-out and annular milling methods. Special care was taken to capture all the layers of interest within the first 70\,nm of the needle without damaging the top layer. The APT data acquisition was performed on a CAMECA LEAP 5000XS instrument. The APT data analysis was conducted using CAMECA’s Integrated Visualization and Analysis Software (IVAS 3.8). We note that nitrogen is systematically undercounted in the AlN barrier due to APT artifacts, arising from preferential field evaporation and the formation of neutral $\mathrm{N_2}$ molecules in III-nitride semiconductors \cite{Tang2015-jm}.

Detailed descriptions and figures on fabrication, measurement setups, and simulations, are provided in the Supplementary Sections.

%\clearpage
\vspace{3mm}\noindent{\bf Competing interests}\\
The authors declare no competing interests.	

\vspace{3mm}\noindent\textbf{Data availability} \\
The data supporting this study are included in the article and its Supplementary Information. Further information are available from the corresponding authors on request. % Source data are provided with this paper.

\vspace{3mm}\noindent{\bf Acknowledgements}\\
We thank Debdeep Jena, Huili Grace Xing, Valla Fatemi, Farhan Rana, Cyrus E. Dreyer, Peter L. McMahon, Anand Ithepalli, Manas Verma for helpful and inspirational discussions. We also thank Yong Sun, Michael Rooks, Lauren McCabe, Yeongjae Shin, Kelly Woods, Sungwoo Sohn for their assistance and guidance in device fabrications. This work was supported by the Air Force Office of Sponsored Research under Grant No. FA9550-23-1-0688 and by the Army Research Office under Grant No. W911NF-24-2-0240. H.X.T. acknowledges support from the Office of Naval Research under Grant No. N00014-23-1-2021. Fabrication facilities use was supported by the Yale Institute for Nanoscience and Quantum Engineering (YINQE) and the Yale SEAS Cleanroom. The TWPA used in this experiment was provided by IARPA and MIT Lincoln Laboratory. The work on STEM made use of the electron microscopy facility of the Cornell Center for Materials Research (CCMR) with support from the National Science Foundation Materials Research Science and Engineering Centers (MRSEC) program (DMR1719875). The Thermo Fisher Spectra 300 X-CFEG was acquired with support from PARADIM, an NSF MIP (DMR-2039380) and Cornell University. N.P. acknowledges support from National Science Foundation Graduate Research Fellowship under Grant No. DGE2139899. P.G. and B.M. acknowledge support from the Air Force Office of Sponsored Research under Grant No. 162023-22608.

\vspace{3mm}\noindent{\bf Contributions}\\
H.X.T. and D.W. conceived the research and designed the experiments. D.W. deposited ALD films and fabricated devices with contributions from M.P. D.W. performed the DC measurement with contributions from C.B. Y.W. constructed RF measurement setup and Y.W., D.W. performed the RF measurement. D.W., C.B. and M.P. provided the theoretical analysis. N.P. did STEM characterization. P.G. did APT characterization. H.X.T., D.M., B.M. supervised the project. D.W., N.P., P.G. and H.X.T. wrote the manuscript with input from all the authors. 

%\section{References}
%\vspace*{-10pt}
%\bibliography{references}

%% file: csupp_body.tex
\setcounter{secnumdepth}{2}  
\renewcommand{\thesection}{\Roman{section}}
%\tableofcontents

\clearpage

\begin{figure*}[b]
    \centering
    \includegraphics [width=1\linewidth]
    {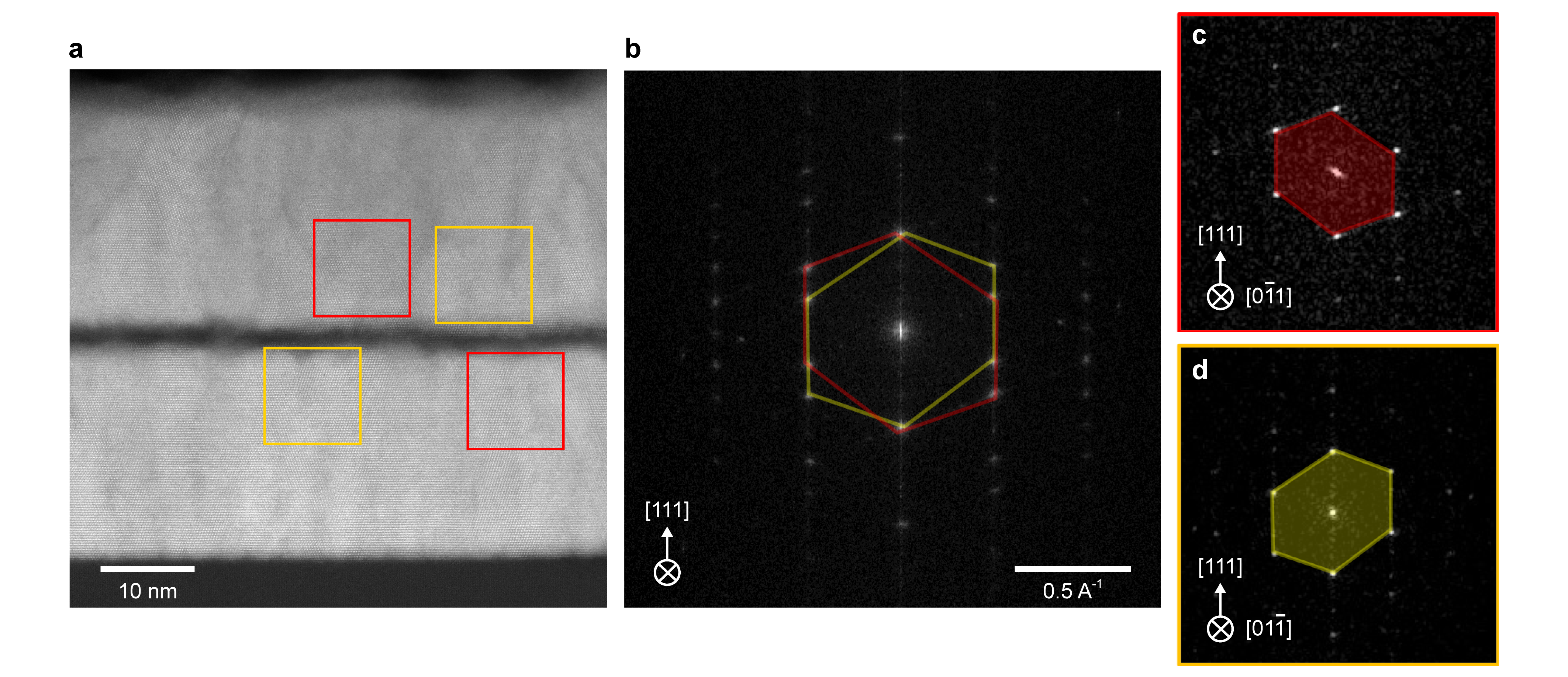}
    \caption{\textbf{STEM image of NbN/AlN/NbN trilayer and corresponding FFTs.} \textbf{a}, Cross-sectional annular dark-field (ADF) STEM image of the trilayer. Regions outlined in red and yellow are selected for local FFT analyses. \textbf{b}, Global FFT of the entire field of view in \textbf{a}, indicating twin domains along the [111] direction. \textbf{c}, FFT of red-marked regions in \textbf{a}, showing grains oriented along the [0$\overline{1}$1] direction. \textbf{d}, FFT of yellow-marked regions in \textbf{a}, showing grains oriented along the [01$\overline{1}$] direction. The shaded regions in \textbf{c,d} highlight the inner diffraction patterns that reflect mirror symmetry along the [111] direction.}
    \label{Fig:S0}
\end{figure*}

\begin{figure*}[t]
    \centering
    \includegraphics [width=1\linewidth]
    {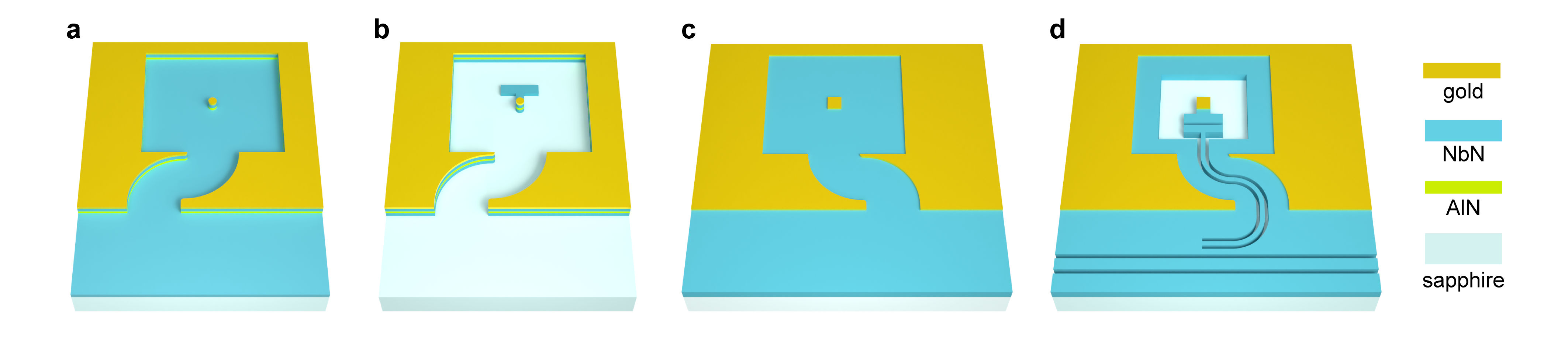}
    \caption{\textbf{Schematics of the fabrication process.} \textbf{a}, Step1 of Q-chip: deposition of gold for bonding; RIE of the top NbN and the AlN barrier for junction formation. \textbf{b}, Step2 of Q-chip: RIE of the bottom NbN to define the shunting pads of qubits. \textbf{c}, Step1 of C-chip: deposition of gold for bonding. \textbf{d}, Step2 of C-chip: RIE of NbN to generate control circuits.}
    \label{Fig:S1}
\end{figure*}

\begin{figure*}[t]
    \centering
    \includegraphics [width=1\linewidth]
    {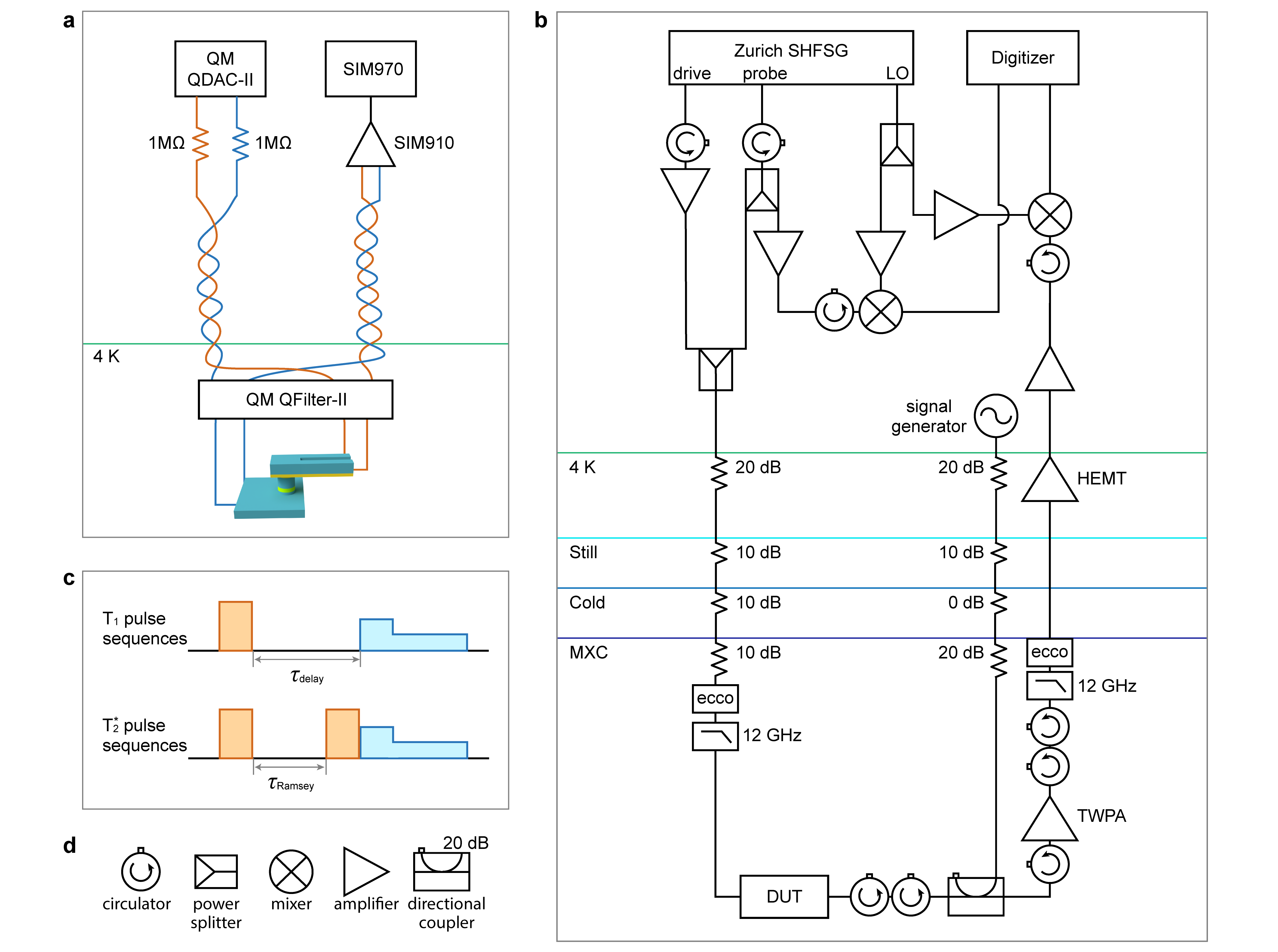}
    \caption{\textbf{Measurement Setup.} \textbf{a}, Setup for IV characterization of junctions. \textbf{b}, Setup for qubit time-domain measurement. Unnamed components in the setup are labeled in \textbf{d}. \textbf{c}, Drive and probe pulse sequences for $T_1$ and $T_2^*$ measurements.}
    \label{Fig:S2}
\end{figure*}

\begin{figure*}[b]
    \centering
    \includegraphics [width=1\linewidth]
    {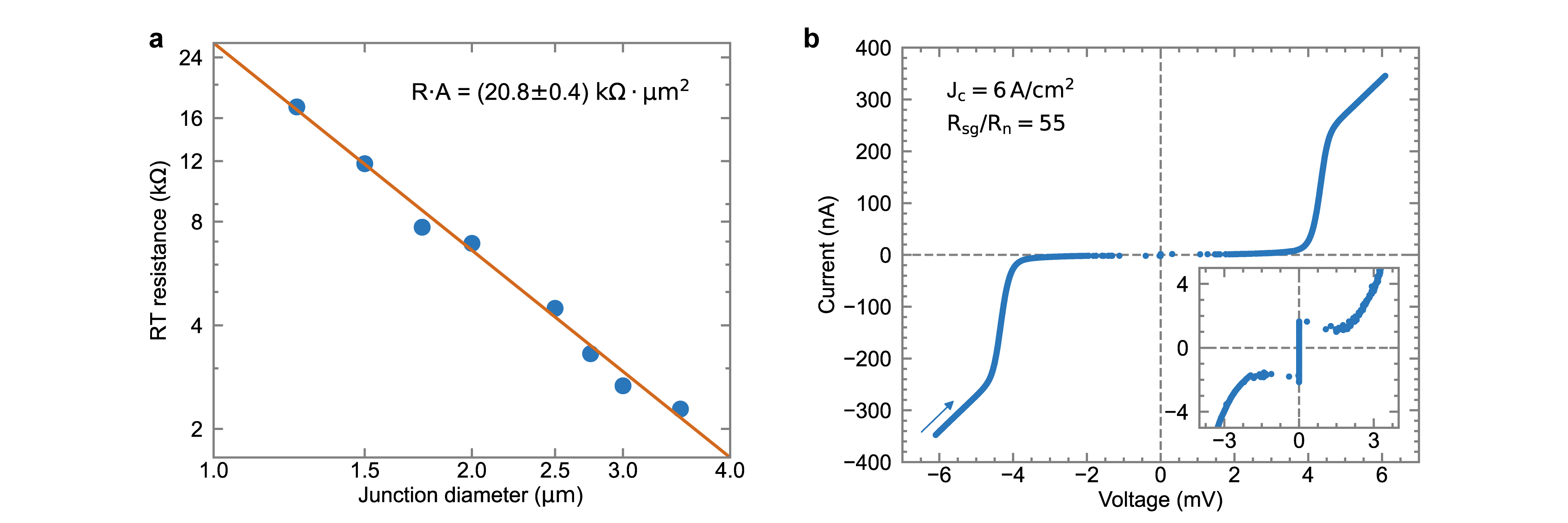}
    \caption{\textbf{DC transport characterization of Josephson junctions.} \textbf{a}, RT resistance of junctions as a function of diameters. The orange line is a fit assuming constant resistance–area product, yielding a value of 20.8\,$\mathrm{k\Omega\cdot\upmu m^2}$ with a standard error of 0.4\,$\mathrm{k\Omega\cdot\upmu m^2}$. \textbf{b}, IV curve of a selected junction, where bias current is swept from negative to positive values. The inset shows a magnified view of the subgap region.}
    \label{Fig:S3}
\end{figure*}

\section{FFT analysis of STEM images}

Fast Fourier transform (FFT) of the cross-sectional STEM image (\figref{\ref{Fig:S0}}{a}) is performed to examine the crystalline orientation across the NbN/AlN/NbN trilayer. As shown in \figref{\ref{Fig:S0}}{b}, the global FFT reveals twin domains, evidenced by two mirror-symmetric diffraction patterns along the [111] direction. To further analyze these domains, multiple regions within the top and bottom NbN layers are selected for local FFTs. The FFT (\figref{\ref{Fig:S0}}{c}) from the red-marked regions in \figref{\ref{Fig:S0}}{a} shows grains oriented along the [0$\overline{1}$1] direction, while the yellow-marked regions produce FFT patterns with orientation along the [01$\overline{1}$] direction (\figref{\ref{Fig:S0}}{d}), confirming the presence of twin domains in the trilayer.

\section{Device Fabrication}

The overall flip-chip fabrication process involves two rounds of electron-beam lithography (EBL) on each chip. Notably, EBL is used based on the authors’ preference; however, the process is fully compatible with photolithography, as the minimum feature size is approximately 0.8\,$\upmu$m. As illustrated in \figref{\ref{Fig:S1}}{a}, the first EBL is performed on the Q-chip to define the trilayer Josephson junctions and the gold bonding regions, using a bilayer resist stack composed of HSQ and PMMA for lift-off. After development, the Q-chip is transferred into an electron-beam evaporation system with ultra-high vacuum conditions ($\mathrm{<10^{-8}}$\,Torr). The $\sim$2\,nm NbO$_x$ is removed by Ar plasma milling, followed by \textit{in situ} deposition of a 10\,nm gold/50\,nm titanium (Ti) bilayer without breaking vacuum. Ti serves as a hard mask for subsequent NbN etching and is later removed using hydrofluoric acid (HF). The top NbN layer and the underlying AlN layer are patterned using reactive ion etching (RIE) with a CF$\mathrm{_4}$/Ar gas mixture; this recipe is consistently applied to all etching steps throughout the process. After being cleaned in acetone and IPA, the Q-chip undergoes the second EBL using PMMA as the resist (\figref{\ref{Fig:S1}}{b}). This step defines the removal of the bottom NbN layer everywhere except for the capacitive pads near the junctions. After solvent cleaning and an HF dip, the Q-chip is ready for bonding.

The C-chip fabrication process follows a similar but simplified procedure. As depicted in \figref{\ref{Fig:S1}}{c}, the 10\,nm-thick gold bonding layer is deposited after EBL patterning and surface treatment. Subsequently, the readout resonators and other control circuits are defined via EBL and formed using the same RIE method (\figref{\ref{Fig:S1}}{d}). Similar to the Q-chip, C-chip is cleaned by organic solvents followed by an HF dip. Prior to bonding, both chips are exposed to a 30\,s, 100\,W Ar plasma treatment to activate the surface. Finally, flip-chip bonding is performed at room temperature using a Tresky T-3000 Pro, achieving alignment precision better than 5\,$\mathrm{\upmu}$m.

\section{Measurement setup}

Devices used for IV characterization presented in Fig.\,2 of the main text are mounted inside a 4\,K cryostat. A QFilter-II from Quantum Machines (QM) is installed near the devices at the 4\,K stage to suppress electromagnetic noise above 65\,kHz. As illustrated in \figref{\ref{Fig:S2}}{a}, the device is current-biased using a voltage source (QDAC-II), with two 1\,$\mathrm{M\Omega}$ resistors in series to ensure stable current flow. The voltage drop across the device is amplified using a voltage preamplifier (SIM910), and the amplified signal is subsequently recorded by a digital voltmeter (SIM970).

Devices used for qubit characterization, shown in Fig.\,3 of the main text, are enclosed in a copper box for electromagnetic shielding and mounted in a Bluefors dilution refrigerator. Frequency-domain measurements, such as two-tone spectroscopy, are performed using a Zurich SHFLI lock-in amplifier, while the time-domain measurement setup is illustrated in \figref{\ref{Fig:S2}}{b}. A Zurich SHFSG signal generator is utilized to provide a qubit drive tone, a probe tone and a local oscillator (LO). The drive and probe tones are combined, attenuated by 50\,dB at cryogenic stages, filtered by an Eccosorb infrared filter and a 12\,GHz low-pass filter, and then sent to the device. The output signal is amplified by a traveling-wave parametric amplifier (TWPA) in the mixing chamber, a high-electron-mobility transistor (HEMT) at 4\,K, and additional room-temperature amplifiers. It is then demodulated using the LO (75\,MHz frequency offset from the probe tone) and recorded by a digitizer (ATS9373). A 75\,MHz reference path, generated by mixing the LO and the probe tone directly from SHFSG, is used for phase correction.

Using the established setup, a qubit is excited by a rectangular $\pi$-pulse, and its $T_1$ is extracted by fitting the probe transmission amplitudes after a set of delay times ($\tau_{\mathrm{delay}}$). As shown by the blue envelope of the readout pulse in \figref{\ref{Fig:S2}}{c}, a short and strong pulse is first applied to the cavity to rapidly populate cavity photons, thereby improving the signal-to-noise ratio \cite{McClure2016-kr}. Similarly, in Ramsey experiments, probe amplitudes are recorded after applying two separated $\pi$/2-pulses to the qubit, and $T_2^*$ is obtained by fitting the exponentially decayed oscillations. During the elevated-temperature measurements (Fig.\,4 of the main text), we employ extensive signal averaging to suppress noise related to the thermal excitation of qubit states: each pulse duration in the Rabi measurements is averaged 250,000 times, and each delay point in the $T_1$ measurements is averaged 600,000 times (50 delay points for $T_1$ extraction).

\section{Junction DC characterization}
NbN/AlN/NbN trilayer junctions are systematically characterized at both room temperature (RT) and 4\,K. To illustrate typical device performance, we present detailed characterization of a representative wafer below. As shown in \figref{\ref{Fig:S3}}{a}, the RT resistance ($R$) scales inversely with the junction area ($A$), following the relation $RA$\,=\,$const$. This behavior confirms a uniform AlN barrier.

\figref{\ref{Fig:S3}}{b} shows the IV characteristic of a representative junction with a 2\,$\mathrm{\upmu m}$ diameter measured at 4\,K. Its superconducting energy gap (2$\Delta$) is 4.3\,meV, and $J_\text{c}$ is approximately 6\,$\mathrm{A/ cm^2}$. The normal-state resistance ($R_\text{n}$) is 14.6\,$\mathrm{k\Omega}$, calculated from the slope of the normal state region, while the subgap resistance ($R_\text{sg}$) reaches 0.80\,$\mathrm{M\Omega}$ at an applied voltage of 3\,mV, as highlighted in the inset. This wafer exhibits the highest $R_\text{sg}/R_\text{n}$ ratio among all wafers characterized, and is selected for qubit fabrication. The low switching current ($I_\text{sw}$) observed in the nanoampere range is likely affected by environmental magnetic fields and residual RF noise, which may be improved by better electromagnetic shielding.

\section{Qubit design and performance}
The $ALDmon$ circuit consists of a transmission line, multiple $ALDmons$, and their associated readout resonators. A distinguishing feature of our design is the flip-chip architecture, with shunting capacitive pads distributed across two separate chips, which simplifies fabrication but adds complexity to simulation. To model the flip-chip structure, electromagnetic simulations are performed using Sonnet, based on a geometry consisting of two substrates separated by a vacuum gap of several tens of nanometers. The resonant frequencies of $ALDmons$ and readout cavities are simulated, incorporating the kinetic inductance of the NbN thin film (7\,pH/square for a 50\,nm-thick NbN layer). Furthermore, the coupling strength ($g/2\mathrm{\pi}$) between an $ALDmon$ and its readout resonator is extracted by intentionally bringing their resonances into degeneracy and analyzing the resulting mode splitting. 

Two wafers, labeled A and B, with $J_\text{c}$ of 6\,$\mathrm{A/cm^2}$ and 0.8\,$\mathrm{A/cm^2}$, respectively (estimated from IV characteristics), are processed to fabricate qubits with transition frequencies ($f_\text{q}$) in the range of 4--5\,GHz. To characterize $ALDmons$' geometry, we introduce the junction energy participation ratio ($p_\text{J}$), defined as the fraction of electric field energy stored in the junction: $p_\text{J}$\,=\,${C_\text{J}}/{C_\mathrm{\Sigma}}$, where $C_\text{J}$ is the junction capacitance and $C_\Sigma$ is the total qubit capacitance. Conventional planar transmons, where large shunting capacitors dominate $C_\Sigma$, exhibit $p_\text{J}$\,$\sim$\,0.02, whereas ``\textit{mergemons}" have $p_\text{J}$\,$\sim$\,0.9, as most of their capacitance originates from the junction itself \cite{Kono2024-lj, Place2021-fg, Merged-qubit}. For wafer A, junctions with diameters of 1\,$\mathrm{\upmu m}$ and 0.8\,$\mathrm{\upmu m}$ are used, resulting in $p_\text{J}$ ranging from 0.30 to 0.20---intermediate between transmons and mergemons. For wafer B, 2\,$\mathrm{\upmu m}$ junctions are employed to realize qubits approaching the mergemon regime. Experimental parameters of the five $ALDmons$ are summarized in Tab.\,\ref{Tab:S1}.

\begin{table}[b]
\centering
\renewcommand{\arraystretch}{1.6}
    \begin{tabular}{|l|l|l|l|l|l|}
        \hline
        \parbox{2.8cm}{\raggedright Qubit} & \parbox{2cm}{\raggedright A1} & \parbox{2cm}{\raggedright A2} & \parbox{2cm}{\raggedright A3} & \parbox{2cm}{\raggedright B1} & \parbox{2cm}{\raggedright B2}\\
        \hline
        $f_\text{q}$ (GHz) & 5.063 & 4.089 & 4.057  & 3.983 & 3.907 \\ 
        \hline
        $E_\text{J}$ (GHz) & {20.020} & {11.790} & {11.545} & {11.980} & {11.189} \\
        \hline
        $E_\text{C}$ (GHz) & {0.172} & {0.196} & {0.197} & {0.182} & {0.188} \\
        \hline
        $g/2\mathrm{\pi}$ (MHz)  & {48.7} & {67.5} & {68.5} & {49.0} & {55.0}\\
        \hline
        $\delta f_\text{c}$ (MHz) & 2.04 & 1.53 & 1.45 & 1.40 & 1.31\\
        \hline
        $f_\text{c}$ (GHz) & 6.8855 & 6.9657 & 7.0874 & 5.8848 & 6.2171 \\
        \hline
        $Q_\text{ci}$ $(\times 10^3)$ & 44.5 & 25.1 & 56.4 & 45.3 & 36.6 \\
        \hline
        $T_1$ $(\upmu\mathrm{s})$ & {$1.43\pm0.03$} & {$2.87\pm0.07$} & {$3.00\pm0.03$} & {$2.66\pm0.05$} & {$3.43\pm0.08$} \\
        \hline
        $T_2^*$ $(\upmu\mathrm{s})$ & {$0.74\pm0.04$} & {$0.65\pm0.04$} & {$1.20\pm0.04$} & {$0.69\pm0.04$} & NA \\
        \hline
        $D_\text{J}$ $(\upmu\mathrm{m})$ & 1 & 0.8 & 0.8 & 2 & 2\\
        \hline
        $p_\text{J}$ & {0.30} & {0.20} & {0.20} & {0.75} & {0.74} \\
        \hline
    \end{tabular}
    \caption{\textbf{Parameters of 5 qubits.} Here, $f_\text{q}$ denotes the qubit transition frequency from the ground to the excited state. $E_\text{J}$ and $E_\text{C}$ are the Josephson and charging energies, respectively. $g/2\mathrm{\pi}$ is the qubit–cavity coupling strength, and the parameters $E_\text{J}$, $E_\text{C}$ and $g/2\pi$ are calculated using scqubits \cite{Groszkowski2021-ng, Chitta_2022}. $\delta f_\text{c}$ represents the shift of the cavity frequency from its bare value due to the qubit coupling. $f_\text{c}$ is the shifted cavity frequency; $Q_\text{ci}$ is the single-photon-level intrinsic quality factor of the shifted cavity. $T_1$ and $T_2^*$ are the qubit’s relaxation time and Ramsey coherence time, respectively. The error represents the standard error from fitting, consistent with those values in the main text. $D_\text{J}$ is the junction diameter, and $p_\text{J}$ is the junction energy participation ratio.}
    \label{Tab:S1}
\end{table}

\begin{table}[t]
    \centering
    \renewcommand{\arraystretch}{1.6}
    \begin{tabular}{|l|l|l|l|l|}
        \hline
        \parbox{2.2cm}{\raggedright $e_{33}$ ($\mathrm{C/m^2}$)} &
        \parbox{2.2cm}{\raggedright $e_{31}$ ($\mathrm{C/m^2}$)} &
        \parbox{2.2cm}{\raggedright $D_\text{J}$ ($\upmu$m)} &
        \parbox{2.2cm}{\raggedright $p_\text{J}$} &
        \parbox{2.2cm}{\raggedright $Q_\text{AlN\_piezo}$}\\
        \hline
        {1.41} & {-0.55} &
        {0.8} & {0.20} & {$8.3\times10^3$} \\
        \hline
        {0.451} & {-0.176} &
        {0.8} & {0.20} & {$8.1\times10^4$} \\
        \hline
        {0.141} & {-0.055} &
        {0.8} & {0.20} & {$8.3\times10^5$} \\
        \hline
        {1.41} & {-0.55} & 
        {2.0} & {0.75} & {$1.9\times 10^3$} \\
        \hline
        {0.451} & {-0.176} &
        {2.0} & {0.75} & {$1.8\times 10^4$} \\
        \hline
        {0.141} & {-0.055} &
        {2.0} & {0.75} & {$1.9\times 10^5$} \\
        \hline
    \end{tabular}
    \caption{\textbf{Simulated $Q_\text{AlN\_piezo}$ for different AlN piezoelectric constants and qubit geometries.} The piezoelectric stress constants $e_{33}$ and $e_{31}$ of row 1 correspond to typical values for wurtzite AlN \cite{coatings14080984}. $D_\text{J}$ is the diameter of Josephson junction in the simulations, and $p_\text{J}$ is the junction energy participation ratio obtained from Tab.\,\ref{Tab:S1}.}
    \label{Tab:S2}
\end{table}

\section{Analysis of loss channels}

We analyze the potential loss channels of the $ALDmon$ devices using a conventional energy participation model and provide promising strategies to further enhance qubit coherence \cite{Wang2015-ch, PhysRevX.13.041005}. Contributions to the overall qubit quality factor ($Q_\text{q}$) can be decomposed into several independent loss channels:
$ \frac{1}{Q_\text{q}}= \frac{1}{Q_\text{JJ\_subgap}}+ \frac{1}{Q_\text{AlN\_piezo}}+ \frac{1}{Q_\text{Au}}+\frac{1}{Q_\text{other}}$. Here, $1/Q_\text{JJ\_subgap}$ denotes dissipation from junction subgap conduction path, $1/Q_\text{AlN\_piezo}$ is the potential piezoelectric loss from the AlN barrier, $1/Q_\text{Au}$ represents loss due to the gold layer, and $1/Q_\text{other}$ includes contributions from surface niobium oxide, the NbN-sapphire interface, crystalline defects in bulk sapphire, and device packaging. Given the unique aspects of our junction materials and the use of gold for flip-chip bonding, we focus our analysis on the first three channels and provide theoretical estimates of $Q_\text{JJ\_subgap}$, $Q_\text{AlN\_piezo}$ and $Q_\text{Au}$.

The subgap conduction path through the AlN barrier is reflected in the finite $R_\text{sg}$. From transport data, a junction with $I_\text{c}$ of 86\,nA exhibits a $R_\text{sg}$ of 260\,$\mathrm{M\Omega}$ at 10\,mK under a 0.5\,mV DC bias. Since $ALDmons$ studied in this work have lower $I_\text{c}$ and require much smaller voltage excitation, this measurement provides a conservative lower bound for $Q_\text{JJ\_subgap}$, estimated as:
${Q_\text{JJ\_subgap}}={\omega_\text{q} C_\mathrm{\Sigma}R_\text{sg}} \gtrsim 5.7\times10^{5}$, which is well above our measured qubit quality factors ($Q_\text{q} = \omega_\text{q}T_1$, ranging from 45$\times10^3$ to 84$\times10^3$, with $\omega_\text{q}=2\pi f_\text{q}$).

Piezoelectric loss can arise when superconducting circuits couple to materials or interfaces with intrinsic piezoelectricity \cite{zhou2024, PhysRevApplied.20.054026}. Wurtzite AlN is a well-established piezoelectric material capable of dissipating electromagnetic energy via electromechanical coupling. However, in our $ALDmon$, the coexistence of cubic and hexagonal phase AlN may weaken the piezoelectricity, and detailed analysis in the following prove that the piezoelectric loss is not the main bottleneck. Since the piezoelectric properties of our AlN film are uncertain, we consider a range of piezoelectric stress constants ($e_{33}$ and $e_{31}$) from typical wurtzite AlN values to one order of magnitude lower. The simulations are performed in COMSOL, and $ {Q_\text{AlN\_piezo}} = \frac{\omega_\text{q}E_\text{em}}{p_\text{J}P_\text{mech}}
$, where $E_\text{em}$ is the stored electromagnetic energy in NbN/AlN/NbN junction and $P_\text{mech}/\omega_\text{q}$ is the dissipated mechanical energy per cycle. As summarized in Tab.\,\ref{Tab:S2}, $Q_\text{AlN\_piezo}$ scales approximately as $1/e^2$, evident from rows 1–3, and also shows strong dependence on qubit geometry, as seen by comparing rows 1 and 4, corresponding to the geometry of qubits A3 and B1 (Tab.\,\ref{Tab:S1}). Experimentally, however, the similar $Q_\text{q}$ values of A3 and B1 confirm that our AlN exhibits very low piezoelectricity, demonstrating that piezoelectric loss is not the dominant qubit loss channel.

Another innovation in our qubit design is the use of gold for flip-chip bonding, including gold on the ground plane for supporting and directly on the junctions for galvanic connection. We first characterized NbN resonators with gold deposited on the ground plane at 3\,K and extracted an equivalent gold resistivity of $6\times10^{-11}\,\Omega\cdot\text{m}$. Based on this low resistive value, the supporting gold layer is deposited far away from qubits, to ensure a minor electric field distribution on the gold and negligible loss contribution. As for gold layers deposited directly above the junctions, experimental data show no clear correlation between $T_1$ and the gold footprint (Tab.\,\ref{Tab:S1}). Theoretically, we estimate the effect of series resistance from gold ($R_\text{Au}$) using a lumped-element model: 
$ Q_\text{Au}\simeq\frac{1}{\omega_\text{q}C_\mathrm{\Sigma}R_\text{Au}}\left(\frac{C_\mathrm{\Sigma}}{C_\mathrm{\Sigma}-C_\text{J}}\right)^2\gtrsim3.3\times10^6 
$, where a gold resistivity of $6\times10^{-9}\,\Omega\cdot\text{m}$ is assumed to provide a highly conservative lower bound for $Q_\text{Au}$. Together, these experimental and theoretical results confirm that the gold layer does not limit device performance.

Based on the above analysis, we conclude that the performance of the devices presented in this work is not limited by the AlN barrier nor the flip-chip bonding approach. The predominant loss mechanism remains under investigation. Future studies will focus on varying $E_\text{J}$ and $E_\text{C}$ of $ALDmons$ to probe possible frequency dependence of $T_1$. Meanwhile, we will pursue improvements through fabrication refinements, including the use of high-quality substrates with advanced surface treatments, minimizing fluorocarbon byproducts during the RIE process, and incorporating additional post-RIE cleaning steps on the NbN surface \cite{kamal2016, Anferov2024-da, hossain2025}. Looking ahead, we anticipate broader applications of $ALDmons$ as coherence times continue to improve.